\documentclass[conference]{IEEEtran}
\usepackage{graphicx,epsfig,color,graphics}
\usepackage{array}
\newtheorem{theorem}{Theorem}
\newtheorem{lemma}{Lemma}
\newtheorem{cor}{Corollary}

\begin{document}

% PAPER TITLE 

\title{Approximate Linear Time ML Decoding on Tail-Biting Trellises in Two Rounds}

% AUTHOR BEGINS HERE 

\author{
  \authorblockN{K. Murali Krishnan , Priti Shankar
      \authorblockA{Dept. of Computer Science and Automation \\
                    Indian Institute of Science, Bangalore - 560012, India\\
                    Email: \{kmurali,priti\}@csa.iisc.ernet.in}}}

\maketitle

% ABSTRACT

\begin{abstract}
A linear time approximate maximum likelihood decoding algorithm on
tail-biting trellises is presented, that requires exactly
two rounds on the trellis. This is an adaptation of an algorithm
proposed earlier with the advantage that it reduces the time
complexity from $O(m\log m)$ to $O(m)$ where $m$ is the number of
nodes in the tail-biting trellis. 
A necessary condition for the output of the algorithm to differ from 
the output of the ideal ML decoder is deduced and simulation results
on an AWGN channel using tail-biting trellises 
for two rate 1/2 convolutional codes with memory 4 and 6 respectively, 
are reported.  
\end{abstract}

\IEEEpeerreviewmaketitle

% MAIN TEXT BEGINS HERE

\section{Introduction}
\label{intro}

Maximum likelihood decoding on tail-biting trellises (TBT) has been 
extensively studied in the literature and 
several linear time approximate algorithms have been proposed,
(see for example, \cite{shu,ma,chep,wang,sundberg}).  Some of these algorithms
may  fail to converge on certain inputs.
Algorithms with guaranteed convergence were studied in \cite{ps4},
but they fail to achieve linear complexity.  In particular, although
the approximate algorithm proposed in \cite{ps4}, achieves
performance close to an ideal ML decoder, it has
a worst case time complexity of $O(m\log m)$, where $m$ is the 
number of nodes in the TBT.  The algorithm exploits the fact that a 
linear tail-biting trellis can be viewed as a coset decomposition
of the group corresponding to the linear code with respect to a specific
subgroup and is an adaptation of the classical $A^{*}$ algorithm. 
The algorithm operates in two phases.  The first phase does a 
Viterbi-like pass on the TBT to obtain certain {\em estimates} which
are used in the second phase to guide the search for the shortest path
corresponding to a codeword in the TBT. 

In this note, the complexity of the approximate
algorithm in \cite{ps4} is reduced to $O(m)$. 
The reduction in complexity is achieved by eliminating the 
use of a heap in the second phase of the original algorithm using the 
well known technique of dynamic programming.  The estimates gathered 
during the first phase are used in the second phase for the 
computation of a {\em metric} for each node in the TBT 
using another simple Viterbi-like pass.  It turns
out that updates performed by the two algorithms are identical for
the shortest path which must be output by the algorithm (although
the metric values computed for other nodes may differ).     

We give an analysis of the algorithm here.  Simulations are included for 
completeness and the two algorithms perform identically as expected.

\section{Background}
\label{background}

A linear tail-biting trellis for an $(n,k)$ linear block code
$C$ over field $F_{q}$ can be constructed as a 
{\em trellis product} \cite{ksch95} of the representation 
of the individual trellises corresponding to the 
$k$ rows of the generator matrix $G$ for $C$ \cite{kv2}. 
The trellis product $T$ of a pair of trellises $T_{1}$ and $T_{2}$ 
will have at $Timeindex(i)$ a set of vertices which is 
the Cartesian product of vertices of $T_{1}$ and $T_{2}$ at that time index, 
with an edge between $Timeindex(i)$ and $Timeindex(i+1)$ 
from $(v_{1},v_{2})$ to $(v_{1}', v_{2}')$, with label $(a+a')$ whenever
$(v_{1},v_{1}')$ and $(v_{2},v_{2}')$ are edges between vertices 
at $Timeindex(i)$ and $Timeindex(i+1)$ in $T_{1}$ and $T_{2}$ 
with labels $a$ and $a'$ respectively for some $a,a'\in F_{q}$,
$0\leq i\leq n-1$, where $+$ denotes addition in $F_{q}$.
Let $\vec g_{i}$, $1\leq i\leq k$  be the 
rows of a generator matrix $G$ for the linear code $C$.
Each vector $\vec{g_{i}}$ generates a one-dimensional subcode of $C$,
which we denote by $C_{i}$. Therefore 
$C=C_{1}+C_{2}+...+C_{k}$, and the
trellis representing $C$ is given by $T = T_1 \times T_2 \times \cdots
\times T_k$, where $T_i$ is the trellis for $\vec{g_i},\ 1 \leq i \leq k$.
%To specify the component trellises in the trellis product above,
%we will need to introduce the notions of linear\cite{ksch95} 
%and circular spans\cite{kv2} and elementary trellises~\cite{ksch95,kv2}.
Given a codeword $\vec{c}=<c_1,c_2,.. c_n> \in C$, the 
{\em linear span} \cite{ksch95} of $\vec{c}$, is the interval 
$[i,j] \in \{1,2,\cdots n\}$ 
which contains all the non-zero positions of $\vec{c}$. 
A {\em circular span} \cite{kv2}
has exactly the same definition with $i>j$. 
  Note that for a given vector, the linear span is unique, 
but circular spans are not.  For a vector 
$\vec{x} = <x_1, \cdots , x_n>$ over the field $F_{q}$, there is a 
unique {\em elementary trellis} \cite{ksch95,kv2} representing 
${\vec{x}}$~\cite{kv2}. This trellis has $q$ vertices at time indices 
$i$ to $(j-1)$ mod $n$, and a single vertex at other positions.
Consequently, $T_{i}$ in the trellis product mentioned earlier, is the 
elementary trellis representing ${\vec{g_i}}$ for some choice of 
span (either linear or circular).  Koetter and Vardy~\cite{kv2} 
have shown that any linear trellis, conventional or
tail-biting can be constructed from a generator matrix whose rows can be
partitioned into two sets, those which have linear span, and those taken to
have circular span. The trellis for the code is formed as a product of 
the elementary trellises corresponding to these rows. 
We will represent such a generator matrix as 
$G = \left[\begin{array}{c} G_l \\ \hline G_c \end{array}\right]$, 
where $G_l$ is the submatrix consisting of rows with linear span, 
and $G_c$ the submatrix of rows with circular span. Let $T_{l}$ denote 
the minimum conventional trellis for the code generated by $G_{l}$.
If $l$ is the number of rows of $G$ with linear span and $c$ 
the number of rows of circular span, the tail-biting trellis 
constructed using the product construction will have $q^c$ start states. 
where, each such start state defines a subtrellis whose codewords form a 
coset of the subcode corresponding to the subtrellis containing the 
all 0 codeword.  

For the description of  the decoding algorithm we assume a 
tail-biting trellis 
with start states $s_0,s_1\ldots s_t$ and final states $f_0,f_1 \ldots f_t$. 
where $t$  is the number of subtrellises. An $(s_i,f_i)$ path is a codeword 
path in trellis $T_i$, whereas an $(s_i,f_j)$ path for $i \neq j$ 
is a non codeword path. For purposes of our discussion we term the 
edge label sequence along such a path as a {\em semi-codeword} as in ~\cite{ps4}.
We assume an AWGN channel with binary antipodal signalling.  When
the edges are given weights corresponding to the log-likelihood
values, ML decoding corresponds to finding the minimum weight 
codeword path in the TBT.

\section{The Two Phase Algorithm}
\label{algorithm}

The algorithm operates in two phases, each taking linear time.
The first phase is a Viterbi pass which computes a function $Cost()$
for each vertex $u$ in the trellis.  This value of cost is used by
the second phase to compute a {\em metric} at each vertex of the trellis.  
The final decoding decision will be based on the {\em metric} values at the 
final nodes of the trellis.

Let $l(u,v)$ denote the length of the shortest path connecting
vertices $u$ and $v$ in the tail-biting trellis.  Note that $l()$ satisfies
the {\em triangular inequality}. ie., $l(u,v)\leq l(u,w)+l(w,v)$ 
for all nodes $u,v,w$ in the trellis.  A {\em codeword} is an
$s_{i}-f_{i}$ path while a {\em semi-codeword} is an $s_{j}-f_{i}$
path, $i,j\in \{1,..t\}$, where $t$ is the number of subtrellises.
Note that all codewords are semi-codewords.  
Define $\delta (u)=min_{1\leq i \leq t}l(s_{i},u)$  
We say an edge $(u,v)\in Section(i)$ if $v\in Timeinde(i)$.
Define the {\em metric at node $u$ for trellis $i$} 
$m_{i}(u)=l(s_{i},u)+\delta(f_{i})-\delta(u)$.  Define {\em metric
at node $u$}, $m(u)=min_{1\leq i\leq t}m_{i}(u)$.   

Suppose  $\delta(u)=x$ and this is the length of an $s_{i}-u$ path,
the first phase of the algorithm assigns the program variable $Cost[u]$
the value $x$ and $SurvTrellis[u]$  the value $i$.  We call the
the $s_{i}-u$ path corresponding to this assignment the {\em survivor} 
at $u$.

These values are used to assign values to the 
program variable $Metric[u]$  in the second phase, which is intended 
to store the value of the metric $m(u)$.  
The trellis corresponding to the minimum 
metric value is stored in the variable $Trellis[u]$.  However, the values assigned to $Metric[ ]$ can be incorrect, in that it is not equal to $m()$. 
The algorithm may even fail to assign a value to $Metric[u]$ for every
node $u$.  We shall derive the conditions under which the algorithm may 
fail to decode correctly.

The program variable $Dist[ ]$ stores the length of the path to the node
corresponding to the minimum value of $Metric[ ]$ in the second phase.  
The program variable $Pred[ ]$ used in both the phases stores the predecessor 
along the paths traced to the node by the algorithm in the respective phases.

The function $Member((u,v),i)$  assumed in the algorithm
description below takes as input an edge $(u,v)$ and integer $i$ and 
returns TRUE if the edge $(u,v)$ belongs to trellis $T_{i}$, FALSE otherwise. 
Note that the function $Member()$ needs only $O(1)$ lookup time 
although the lookup table is of size quadratic on the number of
vertices in the trellis.

\subsection {Phase 1: Estimation} 

\begin{tabbing}

{\bf Initialization:} \\

$for$ \= $each$ $s_{i} \in TimeIndex(0)$\\
\>$Cost[s_{i}]=0$ \\
\>$SurvTrellis[s_{i}] = i$\\
\>$Pred[s_{i}]=s_{i}$\\
$for$ $each$ $v\notin TimeIndex(0)$ $cost[v] = \infty$

\end{tabbing}

\begin{tabbing}

{\bf Estimation:} \\ \\

$for$ $Timeindex := 1$ to $n$ $do$ \\
\hspace{0.3 in} $for$ $each$ $edge$  $(u,v) \in Section(i)$ $do$ \\
\hspace{0.6 in} $Temp = Cost[u] + l[u,v]$\\
\hspace{0.6 in} $if$ ($Cost[v] > Temp$) $then$  \\
\hspace{0.9 in} $Cost[v] = Temp$\\
\hspace{0.9 in} $Pred[v] = u$\\
\hspace{0.9 in} $SurvTrellis[v] = SurvTrellis[u]$\\

\end{tabbing}

Clearly by the end of this phase, $Cost[u]=\delta (u)$
for each vertex $u$ in the trellis. 

Let $j=argmin_{1\leq i\leq t}\delta (f_{i})$.  If the algorithm
assigns $SurvTrellis[f_{j}]=j$, then $survivor$ at $f_{j}$ which 
corresponds to the minimum weight semi-codeword in the trellis 
turns out to be a codeword and the algorithm stops. 
Otherwise, the second phase described below will be executed.

\subsection{Phase 2: Revision}

\begin{tabbing}

{\bf Initialization:} \\ \\

$for$ $each$ $s_{i} \in TimeIndex(0)$ \\
\hspace{0.3 in} $if$ $(Survivor[f_{i}]\ne i)$ $then$
                    $Metric[s_{i}]=\delta (f_{i})$ \\ 
\hspace{0.3 in} $else$ $Metric[s_{i}]=\infty$ /* No processing for $T_{i}$ */\\ 
\hspace{0.3 in} $Pred[s_{i}]=s_{i}$ \\
\hspace{0.3 in} $Trellis[s_{i}] = i$ \\ 
\hspace{0.3 in} $if$ $(Metric[s_{i}]=\infty)$ then $Dist[s_{i}]=\infty$ \\
\hspace{0.3 in} $else$ $Dist[s_{i}]=0$\\
$for$ $each$ $v\notin TimeIndex(0)$ $Metric[v] = \infty$

\end{tabbing}

\begin{tabbing}

{\bf Revision} \\  \\

$for$ $Timeindex := 1$ to $n$ $do$ \\
\hspace{0.3 in} $for$ $each$ $edge$ $(u,v) \in Section(i)$ $do$ \\
\hspace{0.6 in}      $Update(u,v)$ 

\end{tabbing}

\begin{tabbing}
$Update(u,v)$\\
%begin\\
\hspace{0.3 in} $if$ ($not Member((u,v),Trellis[u]$) $return$; \\
\hspace{0.3 in} $temp = Dist[u] + l[u,v] + Cost[f_{Trellis[u]}] - Cost[v]$\\
\hspace{0.3 in} $if$ ($Metric[v] > temp$) $then$ \\
\hspace{0.6 in} $Metric[v] = temp$\\
\hspace{0.6 in} $Pred[v] = u$\\
\hspace{0.6 in} $Trellis[v] = Trellis[u]$\\ 
\hspace{0.6 in} $Dist[v] = Dist[u] + l[u,v]$\\
%end
\end{tabbing}

The second phase attempts to compute the value of the {\em metric},
$m(u)$ for each vertex $u$ of the trellis.  If  the first phase assigned
$SurvTrellis[f_{i}]=i$ for some final node $f_{i}$, for the particular
trellis $T_{i}$ the second phase processing is not required.
We say a Trellis $T_{i}$ {\em participates} in the second phase
if $SurvTrellis[f_{i}]\ne i$ and 
$\delta(f_i)\leq min_{j, SurvTrellis[f_{j}]=j} \delta (f_{j})$. 
The final decoding decision is based 
on the values of the metric at the final nodes of the trellis.
We shall derive the conditions under which the algorithm will 
achieve maximum likelihood decoding on a tail-biting trellis 
for a linear code, when binary antipodal signaling is used over 
an AWGN channel.

\subsection{Final Decision}

If the algorithm does not stop in the first phase,
choose vertex $j = argmin_{1\leq i\leq t}Metric[f_{i}]$.  
The output of the algorithm is the codeword corresponding to the 
$s_{j}-f_{j}$ path obtained by tracing the predecessors of $f_{j}$ till 
$s_{j}$.  The array $Pred()$ stores the predecessors of 
each node along the path the minimizes the value of $metric$.
Note that if $T_{j}$ does not participate in the second phase, 
the path must be traced along $Pred()$ values in the first phase. 

\section{Analysis}
\label{analysis}

For any node $u$, if $Trellis[u]=j$, then 
$Dist[u] \geq l(s_{j},u)$ because the value assigned $Dist[u]$ 
is the length of an $s_{j} - u$ path. Consequently 
$Metric[u]\geq m_{j}(u)$.  We collect these facts into a lemma:

% Lemma 1
\begin{lemma}
During the second phase, if the algorithm assigns for
a node $u$, $Trellis[u]=j$ then $Dist[u]\geq l(s_{j},u)$ and
$Metric[u]\geq m_{j}(u)$.
\end{lemma} 

% Lemma 2
The following simple property of $\delta()$ will be useful:
\begin{lemma}
If $(u,v)$ is an edge in the TBT, the $\delta(v)\leq \delta(u)+l(u,v)$.
\end{lemma}
\begin{proof}
The shortest path from a start node to $v$ cannot be longer than the
shortest path from a start node to $v$ through $u$.
\end{proof}

The following lemma asserts that  the value assigned to $Metric$ by the 
algorithm cannot be smaller than the $Metric$ value of its predecessor node.

% Lemma 3
\begin{lemma}
Let $(u,v)$ be an edge in the Tail-biting Trellis. Let $Trellis[u]=i$ 
Suppose the second phase assigns $Pred[v]=u$ then $Metric[v]\geq Metric[u]$
\end {lemma}
\begin{proof}
An inspection of the algorithm reveals that the algorithm assigns to $Dist[u]$
the cost of some $s_{i}-u$ path.  Hence $Dist[u]\geq l(s_{i},u)$.
By the $Metric$ update rule of the algorithm, $Metric[v]=Dist[u]+l(u,v)+
\delta(f_{i})-\delta(u)$.  Since $Metric[u]=Dist[u]+\delta(f_{i})-\delta(u)$, 
the result follows as $\delta(v)\leq \delta(u)+l(u,v)$ by lemma 2.
\end{proof}

% Corollary 1
\begin{cor}
If the algorithm assigns $Trellis[u]=i$, then 
$Metric[u]\geq Metric[s_{i}]=\delta(f_{i})$
\end{cor}
\begin{proof}
The algorithm initializes $Metric[s_{i}]$ to $\delta(f_{i})$.  
By previous lemma, the value cannot decrease along any $s_{i}-u$ path.
\end{proof}

The algorithm, if assigns any value, must set $Trellis[f_{j}]=j$ and 
$Metric[f_{j}]=Dist[f_{j}] \geq l(s_{j},f_{j}) = m_{j}(f_{j})$ 
for each $j\in \{1,..,t\}$.  Thus, if the shortest path corresponding to a 
codeword in the trellis is an $s_{j} - f_{j}$ path, then if 
$Metric[f_{j}] = l(s_{j},f_{j})$ the algorithm is guaranteed to decode
correctly.  In the following, we derive a condition necessary for the
algorithm to fail.

\begin{theorem}
If the shortest codeword corresponds to an $s_{i}-f_{i}$ path 
$P$, and if $P$ corresponds to the  codeword output by a maximum 
likelihood decoder, then, the two phase algorithm fails to assign 
$Metric[u]=m_{i}(u)$ and $Trellis[u]=i$
for any node $u$ in $P$ {\em only if} there exists $k\ne j\ne i$ 
such that $l(s_{k},f_{j})\leq l(s_{i},f_{i})$.
\end{theorem}

\begin{proof}
Without loss of generality, assume that the all zero codeword was 
transmitted and an ideal ML decoder will output the all zero codeword.  
Again, without loss of generality let 
$P=(s_{1}=)u_{0},u_{1}..u_{n-1},u_{n}(=f_{1})$  
be the shortest $s_{i} - f_{i}$ path in the sub-trellis $T_{1}$
corresponding to the all zero codeword.  We therefore have
$l(s_{1},f_{1})<l(s_{i},f_{i})$ for all $1<i\leq t$.
Let $u$, be the first node along the path $P$ where there exists 
some $1<j\leq t$ such that $m_{j}(u)\leq m_{1}(u)$.  
such node $u$ must exist for otherwise, the algorithm will decode correctly 
as it will assign $Trellis[u_{i}]=1$ with $Metric[u_{i}]=m_{1}(u_{i})$ 
all along the path $P$.  

Note that $m_{1}(f_{1})=l(s_{1},f_{1})$ is 
the value the algorithm would have assigned to $Metric[f_{1}]$ if the
algorithm had assigned $Trellis[u_{i}]=1$ all along the path $P$. 
As the algorithm assigns the minimum value of $Metric$ possible for 
each node, by lemma 3, it must be true that the actual value 
assigned to the $Metric[u]$ by
the algorithm must satisfy $Metric[u]\leq l(s_{1},f_{1})$.  
Since we assume that the algorithm assigned $Trellis[u]=j$,
the value of the metric computed at $u$ must have followed an $s_{j}-u$
path and consequently $Metric[u]\geq Metric[s_{j}]=\delta(f_{j})$ 
(Corollary 1).  Hence $\delta(f_{j})\leq l(s_{1},f_{1})$. 

Now, Assume that the survivor at $f_{j}$ is an $s_{k}-f_{j}$ path, if
$k=j$, we have $l(s_{j},f_{j})\leq l(s_{1},f_{1})$, a contradiction.
Otherwise, the condition stated in theorem holds.

\end{proof}

Now to specialize the above to AWGN channel with binary 
antipodal signaling. The following two results
proved in ~\cite{ps4} are repeated here for completeness.

\begin{lemma}
The space of semi-codewords is a vector space.
\end{lemma}
\begin{proof}
Assume that each of the $c$ vectors in the submatrix $G_c$ of 
the generator matrix is of the form $v_i = [\vec{h_i},\vec{0},\vec{t_i}]$ 
where $\vec{h_i}$ stands for the sequence of symbols before the zero run, 
and is called the {\em head} and $\vec{t_i}$ stands for the sequence of 
symbols following the zero run and is called the {\em tail} and 
$\vec{0}$ is the zero run containing the appropriate number of zeroes. 
Let $\{v_1,v_2 \ldots v_c\}$ be the vectors of $G_c$. Then the matrix 
$G_s$ defined as
$G_s = \left[\begin{array}{c} G_l \\ \hline G_c' \end{array}\right]$, 
where $G_c'$ consists of $2c$ rows of the form 
$[\vec{h_i},\vec{0}],[\vec{0},\vec{t_i}], 1\leq i \leq c$, 
generates the set of labels of all paths from any start node to any 
final node. 
\end{proof}

The following is due to Tendolkar and Hartmann \cite{tendolkar}.

\begin{lemma}
Let $H$ be the parity check matrix of the code and let a codeword 
$\vec{x}$ be transmitted as a signal vector $s(\vec{x})$. Let the 
binary quantization of the received vector 
$\vec{r}= r_1,r_2,\ldots r_n$ be denoted by $\vec{y}$. 
Let $\vec{r'}= (|r_1|,|r_2|,\ldots |r_n|)$ and $S=\vec{y}H^T$. 
Then maximum likelihood decoding is achieved by decoding a received vector 
$\vec{r}$ into the codeword $\vec{y}+\vec{e}$ where $\vec{e}$ is a 
binary vector that satisfies $S= \vec{e}H^T$ and has the property that if 
$\vec{e'}$ is any other binary vector such that 
$S= \vec{e'}H^T$ then $\vec{e}.\vec{r'} < \vec{e'}.r'$ where 
$.$ is the inner product.
\end{lemma}

Combining all the above, we have the following necessary condition
for error.

\begin{theorem}
Assume the $\vec{0}$ codeword is the ML codeword 
corresponding to the path $s_{1}-f_{1}$ in the tail biting trellis.
Let $\vec{y}$ be the binary quantization of the received vector.
Let $r$, $r'$ be as defined in Lemma 4.  For the error pattern 
$\vec{e}$  the two phase algorithm decodes to a vector 
to a vector $\vec{\alpha} \neq \vec{0}$ correspond an $s_{j}-f_{j}$ 
path $j\neq i$ {\em only if} there exists a semi-codeword $C_s$ 
satisfying \[(C_s +\vec{e}) .r' \leq  \vec{e}.r'< (C+ \vec{e}).r'\]
for all nonzero codewords $C$, where the semi-codeword $c_{s}$ either
shares either its head or tail with Trellis $j$.
\end{theorem}

\begin{proof}
Since the ideal ML decoder decodes $\vec{y}$ to $\vec{0}$, we have
$\vec{y}+\vec{e}=0$ or $e=y$.  Let $H$ be the parity check matrix
of the code while $H_{s}$ the parity check matrix for the semi-codeword
vector space established in Lemma 3.  Any binary error vector $\vec{e'}$
which gives the same syndrome as $e$ must belong to the same coset
of the code and hence must have the form  $C+\vec{e}$, where C is a codeword.  
Applying Lemma 5,  we get $\vec{e}.r'< (C+\vec{e}).r'$ for all codewords C,
which proves the right inequality.

To yield the left inequality, first observe that the first phase of the
algorithm does an ML decoding on the semi-codeword space. Any
$s_{k}-f_{j}$ path $P$ in the tail-biting trellis with $k\neq j$ and
$l(s_{k},f_{j})\leq l(s_{1},f_{1})$ corresponds to a semi-codeword
that an ideal ML decoder operating on the space of semi-codewords
will prefer to the all zero codeword.  Hence, by applying Lemma 5 to 
this case and arguing identically as above, we find that for each path such 
$P$ there must exist a semi-codeword $C_{s}$ such that 
$(C_{s}+\vec{e})r'\leq \vec{e}.r'$.  The claim follows as Theorem 1 asserts 
that this condition is necessary for the algorithm to fail to decode
the received vector to the all zero codeword.
\end{proof}

\section{Complexity}
\label{complexity}

Since each phase takes linear time, the algorithm runs in time linear
in the size of the tail-biting trellis. As each pass is Viterbi like,
the worst case number of comparisons performed is bounded by twice
that of the Viterbi algorithm.  The space complexity is 
quadratic in the size of the trellis owing to the lookup table 
of size $t|V|$ required for the $member()$ function, where $|V|$ 
is the number of vertices in the trellis.  However, this is not
a serious drawback as the table can be efficiently implemented
using bit vector representation.

\section{Simulation Results}
\label{simulation}

The results of simulations on an AWGN channel for the two phase algorithm 
are displayed in the figures below.
 The codes used are
a rate $1/2$ memory $6$ convoluational code with a circle size of $48$ 
(same as the (554,744) code convolutional code used in \cite{cav2})
and a rate $1/2$ memory 4 convolutional code with circle size $20$
(same as the (72,62) code used in \cite{cav}).  The performance of the
above codes is compared with that of
the exact ML decoding algorithm in ~\cite{ps4}.  
It is seen that the bit error rate of the 
algorithm approaches that of the ideal ML decoder.

% Figure 3531.eps in IEEE format
\begin{figure}
\centering
\includegraphics[width=3.5in]{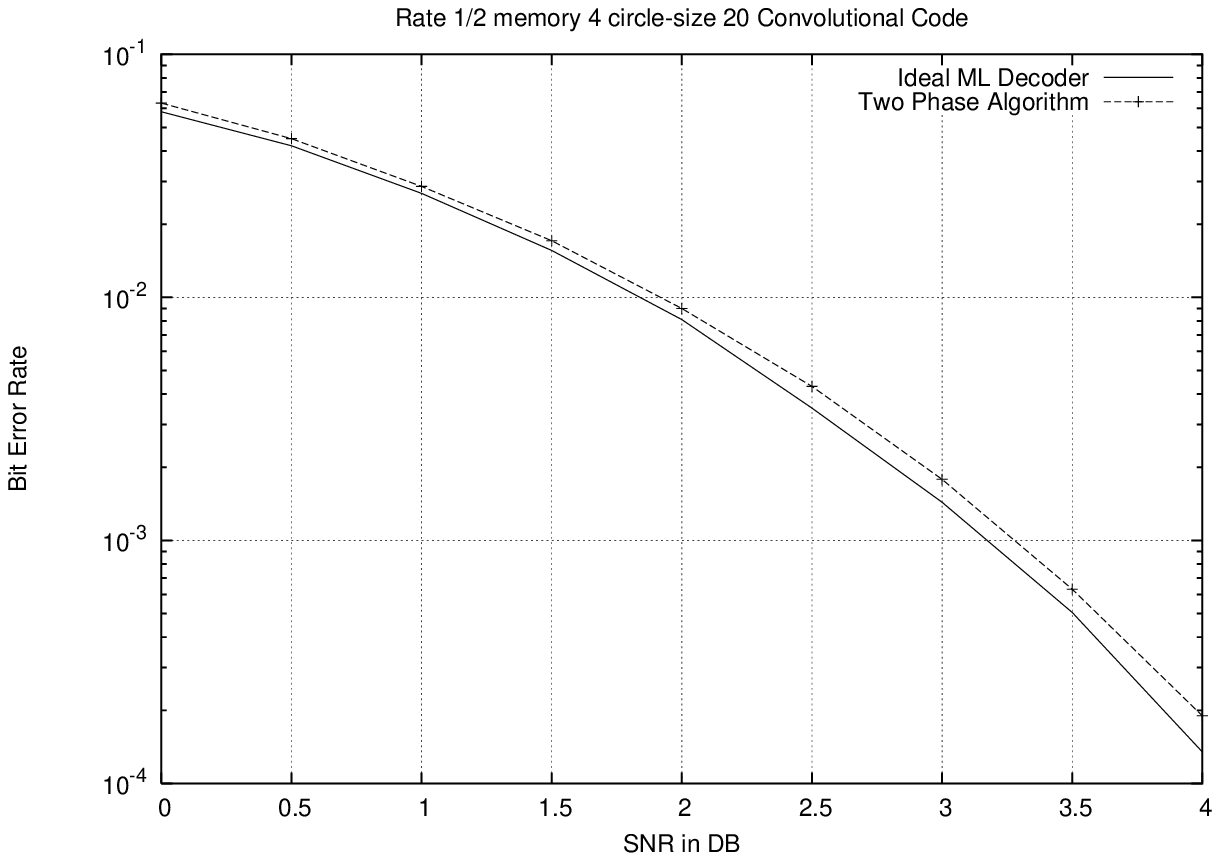}
\label{fig:fig1}
\end{figure}

% Figure 133171.eps in IEEE format
\begin{figure}
\centering
\includegraphics[width=3.5in]{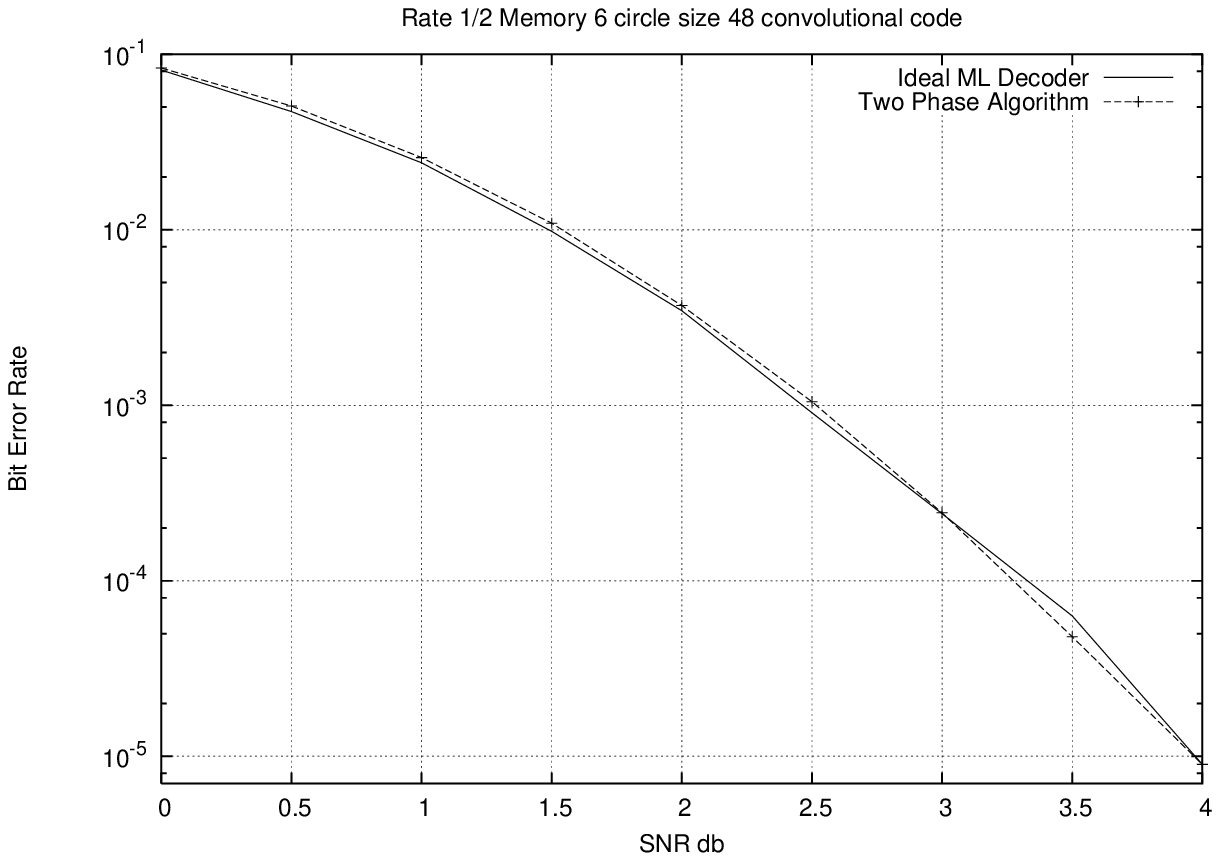}
\label{fig:fig2}
\end{figure}

\section{Discussion and conclusion}
\label{conclusion}

The performance of the algorithm can be improved at the expense of more
storage by tracking more than one paths corresponding lowest values 
of $Metric$ during the second second phase.
However, the time complexity increases 
proportional to the number of stored paths.  Practice has shown that
memorizing the best two paths corresponding to the minimum value of
$Metric$ at each node gives performance almost indistinguishable from
the ideal maximum likelihood decoder \cite{ps4} 

An interesting failure condition of the algorithm is the following:
The algorithm may fail to assign a value to the $Metric$ field for a node
if in the second phase a node fail to belong to any of the 
trellises assigned to the $Trellis$ field of its predecessors by
the algorithm.  If this happens along all paths to all final states,
the algorithm may fail to output a codeword in the second phase. 
Note that the error condition proved handles this case as well.
However this situation never occurred in simulations performed.

From the results of simulations on the rate 1/2, 
memory 4 convolutional code with a circle size of 20 
and a rate 1/2 memory 6 convolutional code with a circle 
size of 48, it is seen that 
the algorithm performs close to the ideal ML decoder.  The performance
is comparable with other linear time approximate methods. The present
algorithm reduces computation to just two Viterbi computation on the
tail-biting trellis and does not require dynamic data
structures like the heap necessary in the orginal versions 
using the A* algorithm \cite{ps4}.

\section{Acknowledgment}
The authors acknowledge Madhu A. S. for implementing the 
algorithm and helping with the simulations.

\end{document}